%% file: usenix2016.tex
\documentclass[letterpaper,twocolumn,10pt]{article}
\usepackage{amsmath}
\usepackage{amsfonts}
\usepackage{amssymb}
\usepackage{graphicx}
\usepackage{tikz}
\usepackage{color}
\usepackage{url}
\usepackage{microtype}
\usepackage{listings}
\usepackage{multirow}
\usepackage{breakurl}
\usepackage{booktabs}
\usepackage{xspace}
\usepackage{algorithm}
\usepackage[noend]{algorithmic}
\def\pgfmathstoreseed#1{\let#1\pgfmath@rnd@z}

\makeatother

\usetikzlibrary{shadows,arrows}
\usetikzlibrary{matrix,positioning}
\pgfdeclarelayer{background}
\pgfdeclarelayer{foreground}
\pgfsetlayers{background,main,foreground}
 
\tikzstyle{materia}=[draw, fill=blue!10, text width=6.0em, text centered,
  minimum height=1.5em,drop shadow]
\tikzstyle{practica} = [materia, text width=8em, minimum width=10em,
  minimum height=3.4em, rounded corners, drop shadow]
\tikzstyle{centro.id} = [materia, text width=8em, minimum width=10em,
  minimum height=3em,  drop shadow]
\tikzstyle{texto} = [above, text width=6em, text centered]
\tikzstyle{linepart} = [draw, thick, color=black!50, -latex', dashed]
\tikzstyle{line} = [draw, thick, color=black!50, -latex']
\tikzstyle{ur}=[draw, text centered, minimum height=0.01em]

\newcommand{\better}{\textcolor[rgb]{00,0.45,0.10}{$\blacktriangle$}}
\newcommand{\worse} {\textcolor[rgb]{0.7,00,00}{$\blacktriangledown$}} 

       \usetikzlibrary{shapes.multipart}
        \usetikzlibrary{shapes.geometric, arrows}
        \usetikzlibrary{positioning,calc}
	\tikzset{my rect/.style={
  rectangle split, 
  rectangle split parts=2, 
  draw,
  rounded corners,
  }
}
	\tikzstyle{process} = [rectangle, minimum width=1cm, minimum height=1cm, text centered, draw=black]
   \tikzstyle{arrow} = [thick,->,>=stealth]

\usepackage{amsthm}

\usepackage{usenix,epsfig,endnotes}
\begin{document}

\date{}

\title{\Large \bf ITect: Scalable Information Theoretic Similarity for Malware Detection}	

\author{
{\rm Sukriti Bhattacharya}\\
University College London
\and
{\rm H\'{e}ctor D. Men\'{e}ndez}\\
University College London
 \and
 {\rm Earl Barr}\\
University College London
 \and
 {\rm David Clark}\\
University College London
} 

\maketitle

\thispagestyle{empty}

\subsection*{Abstract}
\input{abstract}

\input{introduction}

\input{slamm}
\input{ents}
\input{experSetup}

\input{experiments}
\input{countermeasures}
\input{relatedWork}

\input{conclusions}
%
%
\appendix
%
%


{\footnotesize \bibliographystyle{acm}
\bibliography{bibtex}}


\end{document}

%% file: abstract.tex
Malware creators have been getting their way for too long now. String-based similarity measures can leverage ground truth in a scalable way and can operate at a level of abstraction that is difficult to combat from the code level. We introduce ITect, a scalable approach to malware similarity detection based on information theory. ITect targets file entropy patterns in different ways to achieve 100\% precision with 90\% accuracy but it could target 100\% recall instead. It outperforms VirusTotal for precision and accuracy on combined Kaggle and VirusShare malware.

%

%% file: introduction.tex
\section{Introduction}
\label{sec:intro}
 
The number of new variants of malware is increasing at a prodigious rate. These are cheap to manufacture using polymorphic and metamorphic engines. 2014 saw over 317 million \emph{new} malware variants, an increase of 26\% over 2013 \cite{SymISTR20}. Not only is the scale of production increasing but the proliferation of sophisticated techniques is increasing. The proportion of malware able to detect a virtual machine as their execution environment (and change behaviour accordingly) jumped from 18\% to 28\% at the beginning of 2014 \cite{SymISTR20}. 

This combination of increasing volume and sophistication severely threatens the scalability of conventional malware detection methods. Human-performed reverse engineering and analysis does not scale. Dynamic analysis does not scale well (e.g.\xspace time bombs~\cite{CrandallWOSWC06} and red pills~\cite{Paleari:2009}). Static analysis can be defeated (e.g.\xspace limits imposed by opaque predicates~\cite{moser2007limits}). The scalability and accuracy of network neighbourhood analysis methods are promising but when their dependence on specialised information in proprietary databases limits their availability and deployability~\cite{Tamersoy:2014}. 

In this paper, we put forward information theoretic, execution-agnostic, string-based similarity measures as a scalable solution for contemporary malware detection. Similarity must leverage ground truth but by operating directly on strings without any pre-processing, dynamic analysis, reverse engineering, or human intervention, we bypass sophisticated methods of analysis resistance, lower the human cost, and reduce the need for execution in sandboxes or virtual machines. Similarity measures are capable of detecting zero-day malware when it is generated from existing malware.


Information-theoretic similarity
measures, in tandem with machine learning, can accurately, automatically, and efficiently differentiate malware from benign-ware. They identify patterns at a high level of abstraction and difficult to counter by an adversary working at the assembly/source code level (Section \ref{sect:counter}). This has the potential to be a strong move in the malware arms race, inhibiting the current ease of malware production.

We present ITect, which disjunctively integrates two approaches. EnTS (Entropy Times Series) is designed to target polymorphic malware. It extracts a simplified signature from the amplitude and longitudinal variation in a file's entropy and uses this as a machine learning feature. $\mathit{SLaMM}$ (Statistical Language Model for Malware binaries) is designed to target metamorphic malware. It builds $n$-gram language models over zoos of malware and benign-ware. It then uses three different information theoretic metrics, cross entropy, Kullback Liebler divergence, and mean square error, to compare a candidate file with these models.

Surprisingly, both approaches are good at detecting all classes of malware, 
not just the classes for which they were
specifically designed. For the EnTS case, we discovered that metamorphic malware usually
has compressed or encrypted regions.  $\mathit{SLaMM}$ detects polymorphic and packed malware by virtue of the fact that the benign-ware model is a powerful discriminator because benign-ware has low entropy relative to typical malware entropy (Section \ref{sect:experiments}).

In the real world, any classifier must balance precision, recall and accuracy.  
In desktop malware detection, precision is paramount.
In response team malware analysis and classification, recall is paramount.
In our experiments, we deliberately targeted 100\% precision. Our methodology can easily be inverted to target recall instead (or \emph{both} precision and recall, independently). 

We road-tested ITect on a mixture of metamorphic, packed, and polymorphic fresh malware. Some were taken from the Kaggle training sets, some from VirusShare and we used an equal number of fresh benign-ware.  The joint approach achieved 100\% precision and 90.3\% accuracy in detection. This is better than any of the 56 VirusTotal AV engines applied to the same data, the best of which detected the malware with 99.4\% precision and a significantly lower accuracy of 71.4\%. ITect's accuracy and precision remain robust when the proportion of malware in the test set varies from 50\% down to 0\%. As the proportion of malware drops, the accuracy improves and the precision remains the same (Section~\ref{subsect:joining}). ITect's time complexity is linear in the number of files being classified. 

The major contributions of this paper are:
\begin{itemize}
\item We introduce ITect, a new malware detector based on information theoretic similarity measures that operates in linear time.
\item We evaluate ITect and its components with a high level of statistical rigour over a corpus of 30,000 malware and 2,000 benign-ware.
\item We demonstrate that ITect classifies malware with high accuracy and 100\% precision, better accuracy and precision than any VirusTotal AV engine.
\end{itemize}

%% file: slamm.tex
\section{SLaMM}
\label{slamm}

Here, we present SLaMM, a novel malware
classifier that is linear and precision-optimised, two traits essential for our
target use case, the detection of desktop malware.

SLaMM is composed of three subclassifiers; it classifies the suspect $P$ as
malware when its subclassifiers unanimously agree.  Each subclassifiers
compares the similarity of the suspect program $P$ against two program zoos:
$B$, populated with benign-ware, and $M$, with malware.  We build language 
models to model and generalise the information content of these zoos.  For 
instance, these models will capture patterns induced by the differing use of
language constructs between the two zoos, like the frequency of NOP.  Then
we compare these models, and their weighted patterns, against a suspect 
program.

Language models characterise, capture, and exploit regularities in natural
language.  Many language models exist;  one of the simplest and most successful
is the $n$-gram model, which assumes that each word (or symbol) depends
only on the $n-1$ words that precede it.  Under the $n$-gram model, the
probability of the word $w$ occurring at the $i$ location in a sentence is
\begin{equation}
P(w_i~|~h_i) \simeq P(w_i~|~\phi(h_i)) = P(w_i \mid w_{i-n+1}^{i-1}),
\end{equation}
where $h_i$ is $w_i$'s history, the sequence of words that precedes it, and
$\phi$ maps histories into equivalence classes. The Markovian property
distinguishes equivalence classes by the most recent $n-1$ length suffix of
that history, starting from $w_{i-n+1}$ up to $w_i$'s immediate predecessor,
$w_{i-1}$.  

We use maximum
likelihood estimation ($\mathit{MLE}$) to approximate these probabilities.
$\mathit{MLE}$ assigns zero probability to $n$-grams that do not appear in the
training data.  Smoothing takes some probability mass from $n$-grams observed
in the training data and reserves it for unobserved $n$-grams;  we use back-off
smoothing~\cite{Mori97}. 

\begin{table}[t]
\begin{tabular}{cp{0.55\columnwidth}l}
\toprule
$P$ & the suspect program to classify & \\
$q_B$ & \multirow{2}{*}{\raggedright $n$-gram model built from zoo of} 
      & benign-ware \\
$q_M$ & & malware \\
\midrule
$\tilde{P}$ & the histogram of $n$-grams in $P$ & \\
$\tilde{q}_B$ & \multirow{2}{*}{\centering the histogram of $n$-grams in} 
      & $q_B$ \\
$\tilde{q}_M$ & & $q_M$ \\
\bottomrule
\end{tabular}
\caption{The parameters of our classifiers;  below, we treat the variables 
labeled with $\sim$ as probability distributions, under the standard 
construction of a probability mass function from a histogram.}
\label{table:classifier:params}
\end{table}

Table~\ref{table:classifier:params} defines the
parameters of our classifiers.  By construction, the probability mass functions built from each
histogram assume $n$-gram occurrence is independent.
Given the parameters in Table~\ref{table:classifier:params},
SLaMM classifies $P$ as malware
if its three classifiers unanimously classify $P$ as malware.  Formally,  
we have
\begin{align}
\label{slamm}
S(P,q_M,q_B) =&\ C_X(P,q_M,q_B) \wedge C_D(P,q_M,q_B) \\
& \wedge C_\mathit{MSE}(P,q_M,q_B) \nonumber
\end{align}
where we define the classifiers $C_X$, $C_D$, and $C_\mathit{MSE}$ below.

\subsection{Classifiers}
\label{classifiers}

Our classifiers ask ``In which zoo is $P$ more likely to belong?'' using three
different measures on different views on our data.  Our first classifier uses
cross entropy to answer the natural question  ``If our zoo models were program
generators, which is more likely to generate our suspect $P$?''.  Our second
classifier considers only $n$-grams without smoothing.  This has the effect of
making the resulting token distributions formed over $P$ and our two zoos zero
nearly everywhere, concentrating all the probability weight on the observed
tokens.  We then measure their Kullback-Leibler divergence.  Our last 
classifier treats our zoos and $P$ as signals in the unit-less signal space
of probabilities values and repeats the similarity question in this space.
Below, each classifier returns \textbf{T} if it determines $P$ to be malware, 
and \textbf{F} otherwise.  

\paragraph{Cross-Entropy Classifier} Cross-entropy measures the quality of language
models.  Intuitively, it is the Shannon entropy of  $p$ as ``perceived'' by the
model $q$.  
Therefore, a good model is one which has a low entropy, and hence assigns a
high probability to the test data. Formally, cross-entropy is the entropy of a
distribution $p$ as estimated by a model $q$:
\begin{equation}
X(p,q) = -\sum\limits_{w \in W}~p(w)~lg~q(w).
\end{equation}

We are interested in programs, which are sequences of words $w_1^n$ in the
space of programs (i.e. a language), not a single word $w$.  Over a language $L =
\{ w_1^n \mid n \in \mathbb{N} \wedge w_i \in W \}$, the cross entropy of $p$
and $q$ is 
\begin{equation}
\label{xe1}
X(p, q) = - \lim_{n \to \infty} \frac{1}{n} \sum \limits_{w_1^n} ~p(w_1^n)~lg~q(w_1^n).
\end{equation}
For a sufficiently large $n$, we can approximate Equation~\ref{xe1}~\cite{christopher00}:
\begin{equation}
\label{xe2}
X(p, q) \approx  - \frac{1}{n} ~lg~q(w_1^n).
\end{equation}

Given the parameters in Table~\ref{table:classifier:params},
our cross entropy classifier is
\begin{equation}
\label{xe3}
C_{X}(P, q_M, q_B) =  \\ 
  \begin{cases}
    \textbf{T} & \text{if } X(\tilde{P}, q_M)  < X(\tilde{P}, q_B). \\
    \textbf{F} & \text{otherwise}.
  \end{cases}
\end{equation}
%

Consider applying $C_X$ when $P$ is non-obfuscated benign-ware.  The control
flow of $P$, since the triumph of structured programming, is predictable; thus,
control flow constructs generate predictable string patterns.  Malware,
however, obfuscates its control flow to prevent disassembly and to hide itself
behind time bombs.  These obfuscations induce another class of patterns in
malware, viewed as string.  $C_X$ captures these different string patterns,
arising from these different uses of control flow constructs, because
benign-ware's relatively repetitive use of these patterns will surprise $q_M$
and not $q_B$.

\paragraph{Kullback-Leibler Divergence Classifier}

In 1951, Kullback and Leiber~\cite{Kullback51} proposed a new way to compare
two probability distributions associated with the same experiment. This
comparison is now called Kullback-Leibler divergence (KLD) or relative
entropy.   The KLD of the distribution $p$ from the distribution $q$  is 
\begin{equation}
\label{kld}
D(p,q) = ~\sum\limits_{x \in X}~p(x)~lg~\frac{p(x)}{q(x)}.
\end{equation}
KLD quantifies the average number of extra bits
required to encode a distribution $p$ when using the distribution
$q$, which usually models $p$. KLD is not a metric, because 
it is asymmetrical:  $D(p,q) \ne D(q,p)$, in general. 

Given the parameters in Table~\ref{table:classifier:params},
our KL classifier is
\begin{equation} \label{kld1}
C_\mathit{D}(P, q_M, q_B) =  
\begin{cases}
    \mathbf{T} & \text{if } D(\tilde{P},\tilde{q}_M) < D(\tilde{P},\tilde{q}_B) \\
    \mathbf{F} & \text{otherwise}.
\end{cases}
\end{equation}
This classifier returns true when the malware model diverges less from the suspect
$P$ than the benign-ware model does.

Suppose $P$ is metamorphic malware whose code has been
produced by an obfuscation technique introducing NOPs, which is typical junk code introduced by malware engines. $C_D$ cares about the NOP probability mass within $\tilde{P}$ even when NOPs might be captured inside some longer n-grams. If $q_M$ has been generated by malware with NOPs, then the mode of $\tilde{q}_M$ would be closer to NOP n-grams than the mode of $\tilde{q}_B$, making the two probabilities distribution less divergent, according to KLD.

\paragraph{Mean Squared Error Classifier} In statistical modelling, mean
squared error ($\mathit{MSE}$) measures difference between the observed and predicted, or
estimated, values.  In information theoretic terms, $\mathit{MSE}$ measures the
information loss due to the estimator's bias and variance.  It is a metric that
measures the distance between signals.  In our case, we convert $P$ into the
probability distribution $\tilde{P}$, then compare it against each zoo's model
over the signal space of unit-less probability values between $[0..1]$.  Given
the distributions $p$ and $q$, $\mathit{MSE}$ is 
\begin{equation}
\label{mse}
\mathit{MSE}(p, q) = 
    \frac{1}{m} \sum_m (p - q)^2 .
\end{equation}
where $m$ is the number of events in $q$.

Our zoo language models are our estimators and $\tilde{P}$ is set of events we
are estimating.  Given the parameters in Table~\ref{table:classifier:params},
our MSE classifier is 
\begin{equation}
\label{mse1}
C_\mathit{MSE}(P, q_M, q_B) =  
  \begin{cases}
    \mathbf{T} & \text{if } \mathit{MSE}(\tilde{q}_M, \tilde{P}) 
	< \mathit{MSE}(\tilde{q}_B, \tilde{P}) \\
    \mathbf{F} & \text{otherwise}.
  \end{cases}
\end{equation}
Our MSE classifier returns true when $P$ is closer to $q_M$ than $q_B$ in the 
signal space of probability values.

$C_{MSE}$ captures unusual identical patterns better than other two classifiers. An example of these patterns is opaque predicates. Ming et al. showed that opaque predicates are frequent in current malware~\cite{ming2015loop}.  Blackhats could use Obfuscator-LLVM \cite{junod2015obfuscator} to introduce opaque predicates in the malware. Then, the opaque predicates captured by $\tilde{P}$ will be closer to those captured by $\tilde{q}_M$, under $\mathit{MSE}$.

%% file: ents.tex
\section{EnTS: Entropy Time Series Analysis}

Polymorphism is a common method of disguising malware and resisting analysis.
Polymorphism hides parts of a program through encryption or packing. During execution, the relevant parts are unpacked or decrypted at need. One side effect is variation in the entropy between different regions of the file, creating a signal that can be used to distinguish malware. Time series have been widely studied in the literature \cite{brockwell2013time}, applied in many different fields, and have often been used for prediction. Two main approaches to their use can be distinguished: analysis of the series trend to estimate the next value \cite{chatfield2000time}, and grouping of time-series by similarity \cite{liao2005clustering}. In what follows we use the second approach to design the Entropy Time Series or EnTS, an analysis on bit strings that targets polymorphic malware in a scalable way.

\paragraph{EnTS Design}
We summarise the overall approach then consider some details of the algorithm. 
We consider files taken from two zoos, one of benign programs and the other of malware. We consider each file as a stream of chunks (fixed length segments) each with an associated entropy value. The entropy of each chunk is calculated from the byte frequencies for each chunk from which a probability distribution on the bytes is calculated. We then calculate the entropy profile, or time series, associated with each file in the following steps.
\begin{enumerate}
\item Use the smaller median length of file for the two zoos to set a fixed number of chunks, $N$, to be chosen from each file.
\item Use a (Procrustean) deterministic algorithm to choose evenly spread chunks from each file to produce a vector of $N$ chunks in order.
\item Calculate the entropy for each chunk in the vector to obtain an $N$-vector of entropies. 
\item Apply a wavelet transform to obtain an $N$-vector of smoothed entropies with less trivial variation. 
\end{enumerate}
This $N$-vector of smoothed entropies forms the time series for each file. We can then interpret each time series as a coordinate in an $N$-dimensional space and train a machine learning classifier to distinguish malware and benign-ware. From now on we will use \emph{ML classifier} for the phrase \emph{machine learning classifier} to avoid confusion with general classifiers.

%

We discuss aspects of the algorithm in more detail. The entropy profile of a file, $F$, will be computed as a Discrete Haar wavelet Transformation (see equation \ref{eq:dwt}). The Haar wavelet requires that the number of scaling factors, $N$, must be a power of 2, i.e. $N = 2^\alpha$ for some $\alpha \in \mathbb{N}$. 
\[\alpha=\lceil\log (min(median(Z|_M),median(Z|_B))/c)\rceil\]

where $Z|_B$ is the benign zoo, $Z|_M$ is the malware zoo and $c$ is the \textit{chunk size} (fixed at the beginning, see section \ref{sec:ents:param}). The ceiling operator produces an integer between the two median lengths for the two zoos. Then, for each program $P$ in the zoos, we construct its entropy profile, $F$. We first divide the file into chunks of size $c$. 

Once we have the chunk division for a file, we need to reduce the number of chunks to $N$, in order to fit the mother wavelet, defined by:
\begin{equation}
W(N,b) = \frac{1}{|N|^{1/2}} \sum_{j=1}^{|C|} H(C_j) \cdot \Psi_{HAAR} \left( \frac{t_j - b}{N} \right).
\label{eq:dwt}
\end{equation}
where $N$ corresponds with the dimensions of the final $N$-vector space, $b$ is a shifting parameter, $H(C_j)$ are the entropy values, $|C|$ is the total number of chunks, $t_j$ is the current chunk $j$ in the sequence and $\Psi_{HAAR}(t)$ is the Haar wavelet defined by:
\begin{equation}
\Psi_{HAAR}(t) = \left\{ \begin{array}{rr}
1 ,& 0 \leq t < 1/2\\
-1, & 1/2 \leq t < 1\\
0, & \mathrm{otherwise}\\ 
\end{array} \right.
\label{eq:haar}
\end{equation}
The Haar wavelet is chosen because it approximates a step function from the original function. EnTS focuses on the variation patterns, therefore a smooth step function provides all the information it needs about the most relevant entropy variations. 

The selection process of the coefficients extracts equidistant coefficients from the profile. The first and last chunks have special status because file head and tail are usually relevant parts in malware analysis. To choose the rest of the coefficients, we calculate an increment value $inc = (|C|-1)/(N-1)$ to get the next chunk index using the floor of the accumulation of this factor as the next chosen index.
For each chosen chunk we calculate its Shannon entropy on the basis of the byte frequencies of the chunk:

$$H(C_j) = -\sum_{b \in C_j} p(b) \log_2 p(b),$$

where $p(b)$ is the probability of byte $b$ within the $j$th chunk, $C_j$, of program $P$, calculated from its frequency count within the chunk. 

Then, we calculate the discrete Haar wavelet transformation. Each iteration in the process is divided into two parts: calculating the scale coefficients and calculating the detail coefficients. The scale coefficients contain the most relevant information about the signal while the detail coefficients contain information about the small variations. In each iteration, the coefficients used are the scale coefficients for the previous iteration, e.g. in iteration number 2 only the scale coefficients of iteration 1 are used to calculate the scale and detail coefficients of iteration 2, and the other wavelet coefficients are not modified. According to the Haar wavelet equations, a scale coefficient is calculated by:
$$s^1_i = \frac{1}{\sqrt{2}} (x_i + x_{i+1}), \ \ s^\alpha_i = \frac{1}{\sqrt{2}} (s^{\alpha-1}_i + s^{\alpha-1}_{i+1}), \ \alpha>1,
$$
and a detail coefficient is calculated by equations:
$$d^1_i = \frac{1}{\sqrt{2}} (x_i - x_{i+1}), \ \  d^\alpha_i = \frac{1}{\sqrt{2}} (s^{\alpha-1}_i - s^{\alpha-1}_{i+1}),\ \alpha>1,
$$ 
The scale coefficients are positioned at the beginning of the wavelet and the detail coefficients after the scale coefficients. For example, with $\alpha = 3$, the iterations generate the coefficients as follows:

$$\begin{array}{c}
(x_0,x_1,x_2,x_3,x_4,x_5,x_6,x_7)\\
\downarrow \uparrow\\
(s^1_0,s^1_1,s^1_2,s^1_3,d^1_0,d^1_1,d^1_2,d^1_3)\\
\downarrow \uparrow\\
(s^2_0,s^2_1,d^2_0,d^2_1,d^1_0,d^1_1,d^1_2,d^1_3)\\
\downarrow \uparrow\\
(s^3_0,d^3_0,d^2_0,d^2_1,d^1_0,d^1_1,d^1_2,d^1_3)\\
\end{array}$$

In the final iteration the wavelet, $W$, has been constructed. We can use it to reduce the noise from the entropy time series, using a \textit{threshold}, $\tau$, on the wavelet coefficients in this final iteration. Those values that are below the threshold are set to 0. This process improves the performance of the classification task by eliminating minor variations in the original profile.

Lastly, we apply the inverse wavelet transformation to reconstruct  the entropy profile without the noise.

$$x_i = \frac{1}{\sqrt{2}} (s^1_k + d^1_k), \ \ s^\alpha_k = \frac{1}{\sqrt{2}}(s^{\alpha+1}_k + d_k^{\alpha + 1}), \ \alpha >0$$
$$x_{i+1} = \frac{1}{\sqrt{2}} (s^1_k - d^1_k),  \ \ s^\alpha_{k+1} = \frac{1}{\sqrt{2}}(s^{\alpha+1}_k - d_k^{\alpha + 1}), \ \alpha >0$$

The resulting coefficients vary between 0 and 8 because of the choice of chunk size and will be used as coordinates of the entropy time series in the classification space. This space allows the creation of scalable models based on ml classifiers and significantly improves the speed of the classification process. The ml classifier will infer a way of discriminating the zoos, focused on targeting 100\% precision, that is one of our main goals.

\paragraph{Chunk Size} 
 \label{sec:ents:param}  
 Chunk size is a critical parameter for EnTS. Chunks are file segments but we also considered sliding windows as an alternative. This was quickly rejected because it would add more longitudinal variation into the entropy profile signal, not less as is the aim. 
 
 Given that the the atomic constituents of chunks are bytes, it is easy to see that a chunk size of 256 bytes is optimal with respect to the amplitude of entropy variation. There are $256= 2^8$ possible different bytes. Entropy of a chunk will be maximal when every possible byte has equal probability, so the minimum chunk size that allows the maximal possible variation in entropy (from 0 to 8 bits) is 256. On the other hand we want as many chunks as possible in each file so we also want the length of chunks to be as small as possible.

\paragraph{Example}

Consider a zoo of just two binary files, $P$ and $Q$, and a chunk size of $c$. These programs, considered as binary strings, are divided in chunks. Suppose that $length(P) = 20c$ and $length(Q) = 6c$. Each chunk is related to a wavelet coefficient, therefore, the number of coefficients would be 20 for $P$ and 6 for $Q$. However, the Haar wavelet requires $2^\alpha$ coefficients. Suppose that we choose $\alpha = 3$, then we need $N=2^\alpha=8$ coefficients. For $P$ we need to contract the number of chunks from 20 to 8 and for $Q$ we need to increase the number of chunks from 6 to 8. In order to choose these chunks, we generate a subset of the current chunks using a jump factor for each file. The chunk index is initially set to 0, and it is incremented in every step by $inc_1=19/7=2.71$ for $P$ and $inc_2 = 5/7 = 0.71$ for $Q$. The indices are selected using the floor of the accumulated jump value, so the chosen indices will be:
$$ I_P = (0,2,5,8,10,13,16,19) \quad I_Q = (0,0,1,2,2,3,4,5)$$
Now, for purposes of illustration, we focus on $P$. The entropy of each chunk is calculated, defining an N-vector of entropy values for each file which is considered as an entropy time series. In order to remove noise and simplify each time series by obtaining the reconstruction coefficients, we apply the discrete Haar wavelet transformation. Assume that the entropy values for $P$ are $(4,5,4,1,1,2,1,2)$ then the wavelet transformation process will give us:
$$\begin{array}{l}
W_P|_{\alpha=1} = (6.4,3.5,2.1,2.1 \ | \ -0.7,2.1,-0.7,-0.7)\\
W_P|_{\alpha=2} = (7,3 \ |\ 2,0,-0.7,2.1,-0.7,-0.7)\\
W_P = (7 \ | \ 2.8,2,0,-0.7,2.1,-0.7,-0.7)\\
\end{array}$$
We apply the threshold, in this example it is 0.75, to $W_P$ and we get $W_P = (7,2.8,
2,0,0,2.1,0,0)$. Then, we apply the reconstruction process to $W_P$ and we get the reconstructed signal as $(4.5,4.5,4,1,1.5,1.5,1.5,1.5)$. These values are the coordinates of $P$'s profile in the space.

%% file: experSetup.tex
\section{Experimental Setup}
\label{sec:ExpSetup}

We designed our study to validate the accuracy and scalability of EnTS, then
SLaMM.  Our classifiers are linear;  our study demonstrates their scalability.
Finally, we explore how ITect combines EnTS and SLaMM, improving their accuracy
while achieving perfect precision over our data sets.  

EnTS was designed to detect polymorphic malware. How good 
it is at detecting all types of malware? \textbf{RQ1}: \textit{Does EnTS accurately and precisely detect
malware?} 

Following related work, we consider a detector to be accurate when its accuracy
is at least 90\% and precise when its precision is 100\%.  EnTS uses a ML
classifier. To determine how much of its performance is due to its ML
classifier and how much to its similarity metric, we compared EnTS with other
information theory similiarity measures, using the same parameters and ML
classifier and ask \textbf{RQ1.2}: \textit{How does EnTS' accuracy and
precision compare to that of other information theory similarity measures, like
NCD?}

We designed SLaMM to detect metamorphic malware. How well does it detect other
types of malware? \textbf{RQ2}: \textit{Does SLaMM accurately and precisely
detect malware?} The scalability of our classifiers is a key contribution of
this work.  We explore our scalability by asking \textbf{RQ3}: \textit{Do EnTS
and SLaMM scale better than NCD, CR and SE?} ITect combines SLaMM and EnTS.
Does it outperform them? \textbf{RQ4}: \textit{How accurately and precisely
does ITect detect malware?} Finally, we ask \textbf{RQ5}: \textit{Can ITect
improve the results of the AV engines with 100\% precision?}

\paragraph{Data Collection}
\label{sec:ExpSep:data}

We evaluate our classifiers against different malware concealment
strategies, in particular polymorphism (based on compression and encryption)
and metamorphism. The datasets used have been extracted from public
repositories.

The first dataset is the \textbf{Kaggle malware competition dataset}\footnote{\url{https://www.kaggle.com/c/malware-classification}}. It contains two subsets: train and test. Kaggle's test subset is not labelled, so we train and test on the train subset. It is composed of 10,869 Malware files.  The dataset contains 9 malware families whose features are summarised in Table \ref{tab:classes}. There are two files per malware: a byte representation (hexdump) and an asm file with IDA Pro information from the disassembly process. We used \texttt{xdd}\footnote{\url{http://linux.about.com/library/cmd/blcmdl1_xxd.htm}} to convert the hexdumps to binary executables. This dataset was published February 2015.

We collected \textbf{packed malware from VirusShare}\footnote{\url{http://virusshare.com}}. We desired Win32 malware whose packing system was known, so we focused on the first malware set of VirusShare, uploaded in June 2012. This database is composed of approximately 112,000 Malware files. By combining Yara\footnote{\url{http://yara.readthedocs.org}} with packer rules extracted from the YaraRules project\footnote{\url{http://yararules.com/}} and information from VirusTotal\footnote{\url{https://www.virustotal.com/}} we extracted 3,000 Win32 malware with known packers. Around 70 specific packing systems were detected by Yara, however, several of them came from the same family, so we focused on the most frequently occurring families (see Table \ref{tab:packers}). 

\begin{table}[t]
\centering
\footnotesize
\begin{tabular}{c|rrrcc}
Class & Instances & EnTS & SLaMM & Conc. &Type\\
\hline
Ramnit & 1541 & 1418&400 & Poly & Worm\\
Lollipop & 2478 & 2279& 400& Poly & Adware\\
Kelihos\_3 & 2942 & 2716&400& Poly & Botnet\\
Vundo & 475 &440&475& Meta &Trojan\\
Simda & 42 & 37& 0& Poly & Botnet\\
Tracur & 751 &  687& 400&Poly & Trojan\\
Kelihos\_1 & 398 & 360& 0&Encr &Botnet\\
Obf.ACY & 1228 & 1127& 1228&Meta &Trojan\\
Gatak & 1013 & 936& 400&Poly & Trojan\\
\hline
Total & 10868 &10000 &3703\\
\end{tabular}
\caption{Information about Kaggle classes:number of instances, instances used in EnTS and SLaMM's experiments, concealment strategies and types of malware.}
\label{tab:classes}
\end{table}

We synthesised a \textbf{mixed dataset} by sampling: roughly 1/3 from Polymorphic data (Kaggle), 1/3 from Metamorphic (Kaggle) and 1/3 from Packed. The resulting dataset has 2,000 malware instances. In industry, white hats often must analyse different kinds of malware at the same time. This dataset aims to emulate this scenario. 
Finally, we collected \textbf{2,000 benign files from Windows OS}. 

To generate data for EnTS, we uniformly sampled 10,000 files from the Kaggle
dataset without replacement and randomly divided it into five equal sets. We
uniformly divided the (packed) files from the VirusShare dataset into three
groups of 1,000 (See Tables~\ref{tab:classes} and \ref{tab:packers}).

SLaMM trains different language models for each category of malware
(polymorphic, metamorphic and packed) and one for benign-ware, then compares
any candidate file with each of the four models. So for SLaMM we chose
different malware sets from files labelled metamorphic, polymorphic and packed
(see Tables \ref{tab:classes} and \ref{tab:packers}). 

\begin{table}[t]
\centering
\footnotesize
\begin{tabular}{c|rrrr}
Packer & Samples 1 & Samples 2 & Samples 3 & SLaMM\\
\hline
Armadillo & 37 & 41& 42 & 78 \\
ASPack & 58 & 64& 58 & 122\\
ASProtect & 16 & 19& 21 &35 \\
Borland & 261 &251& 229 &512\\
NET & 57 & 88& 106 &145\\
PECompact & 47 & 46& 47 &93 \\
UPX & 268 & 266& 265 &534\\
Rest & 256 & 225& 232 & 481 \\
\end{tabular}
\caption{Information about VirusShare packers: number of instances in EnTS samples, and instances in SLaMM experiments.}
\label{tab:packers}
\end{table}

\paragraph{Algorithms and Parameters}
\label{sec:ExpSetup:algs}

The evaluation consists of four main steps. The first is data selection. In
every experiment we have chosen the same number of malware and benign-ware
instances. The training data consists of two thirds of the instances and the
remaining third comprises the test data. The test data is always fresh data for
either approach and is randomly selected by uniform sampling at the beginning
of the process. The second step consists of search space generation for
classification in EnTS or n-gram counting for construction the language models.
Once these are prepared, we train a classifier or construct a language model as
appropriate. Finally, we evaluate malware detection on the test set, recording
the accuracy and the false positive rate. 

\paragraph{EnTS} EnTS applies machine learning to entropy profiles.  To compare
against EnTS, we implemented three other information theoretic features of
binary strings from the literature. The first is the \textbf{compression rate}
(CR) which calculates the ratio of the compressed length to the uncompressed
length for a given file compressor and  is related to the Kolmogorov complexity
of the specific file~\cite{li2013introduction}.  We chose LZMA2 as the
compressor and its maximum compression parameters and the maximum windows size,
i.e., 4GB, using the package 7zip .

The second is the \textbf{Normalised Compression Distance} (NCD), which
approximates the Normalised Information Distance~\cite{li2013introduction}, a
universal, generic, information theoretic metric. Formally, NCD is 
\begin{align*}
NCD(P,Q) = \frac{C(PQ) - min\{C(P),C(Q)\}}{max\{C(P),C(Q)\}},
\end{align*}
where $P,Q$ are strings, $PQ$ is their concatenation, and $C(\cdot)$ is the
compressed size function for a specified compressor. NCD also uses LZMA2 as the
compressor. 


Finally, we compare against \textbf{Structural Entropy (SE)}.
Sorokin introduced this technique in 2011~\cite{sorokin2011comparing} and 
Baysa et al.
applied it to metamorphic malware
in 2013~\cite{mark2013}. It divides a file into chunks,
calculates the entropy of each chunk, then groups the chunks into
arbitrarily sized segments (the information for each segment is its average entropy
and its size). It generates a similarity matrix, performing a pairwise comparison
on the files based on Levenshtein distance. This approach is
$O(n^2)$, where $n$ is
the number of files. Further, the variable number and variable size of segments in a file means 
this approach may determine a file with more segments to be 
totally different from another file with fewer segments even though the overall entropy 
pattern in the two files is similar.
 EnTS escapes this problem: it 
extracts an fixed length entropy time series from a file as a token stream and 
operates directly on this time series and therefore all of the file's information 
at once.
The parameters chosen for this comparison are the same
as those used in both Baysa and Sorokin's work: $\tau=0.3$, $c_\epsilon=0.6$
and $c_\alpha=1.4$.

The parameters for \textbf{EnTS} are: the chunk size is 256, $N=512$ elements ($2^9$) because the smaller zoo (packed files) has an average size of 116 KB, the threshold is 0.5, and $\alpha = \lceil \log (116 \cdot 2^{10} / 2^{8}) \rceil = 9$.

The algorithms used for classification have been taken from the classification
literature. In order to improve the learning process, we have used a
multiple-learners approach, where different classifiers are combined in order
to divide the learning process, specialising different regions of the space.
The multiple learning approach used here is \textbf{Random
Forest}\footnote{\url{https://cran.r-project.org/web/packages/party/index.html}}
with 100 trees. This method uses voting to combine different classifiers.
The base classifiers chosen are \textbf{Inference
Trees}\footnote{\url{https://cran.r-project.org/web/packages/party/index.html}}
classifiers. These are useful for incrementing the accuracy and increasing the
precision, because they penalise false positives during
construction~\cite{hothorn2004bagging}. By combining these two
methods, we construct, as the results show, a competitive 
classifier/detector in each case
that optimally finds the boundary between malware and benign-ware for
each measure.

\paragraph{SLaMM} For SLaMM, we train four different n-gram language models
(polymorphic, metamorphic, packed, and benign). The
only parameter is the size of the $n$-gram, i.e. $n$.  We chose a tri-gram (n=3)
language model for each corpus to strike a balance between memory use and
providing sufficient information to produce good results. To extract $n$-grams
from our zoos and suspect programs, we cut each program's hexdump into
overlapping substrings of length $n$.

%% file: experiments.tex
\section{Experimental Analysis}
\label{sect:experiments}

This section starts evaluating EnTS, follows with SLaMM and analyse their scalability. It concludes evaluating the ITect, as the combination of both. 
  
%
 

\paragraph{EnTS Evaluation}
\label{sec:exps:entsEv}

NCD and SE generate a similarity matrix while EnTS and CR describe point coordinates for the profiles and the compression rate, respectively. Applying machine learning to EnTS and CR is straightforward because we have the points and we only need to discriminate them. For NCD and SE we consider each row of the training similarity matrix as coordinates, due to the number of files is fixed. This provides the points that the machine learning algorithm uses. For testing, we will consider the coordinates as the similarities among the test files and the training files. The machine learning algorithms chosen are non-deterministic approaches (they choose a random seed during the initialization process), then, we need to generate different models to measure their median performance \cite{hothorn2004bagging}. Hence, each experiment has been carried out 100 times, and the median and standard deviation has been provided to compare the results. Furthermore, in order to compare different algorithms, we have applied the Wilcoxon test to evaluate whether there is statistical significance among the results or not. We consider that there is statistical significance when the $p$ value is less than 0.05 using EnTS as benchmark. In order to reduce the redundancy of correlated variables in the space, we have eliminated those dimensions whose Pearson correlation was higher than 0.8 with respect to other dimension. This reduces the space to the 5\% of the original dimension.

\begin{table}
\centering
\footnotesize

\begin{tabular}{l|rrrr}
\multicolumn{1}{c}{Data} &\multicolumn{1}{c}{NCD}&\multicolumn{1}{c}{CR}&\multicolumn{1}{c}{SE}&\multicolumn{1}{c}{EnTS}\\
\hline
Kag 1&\worse \textit{93.7 $\pm$ 0.5} & \worse 78.5 $\pm$ 0.5 &\worse 87.3 $\pm$ 0.6 & \textbf{95.1 $\pm$ 0.2}\\
Kag 2&\worse \textit{92.1 $\pm$ 0.6} &\worse 80.0 $\pm$ 0.2 & \worse 88.9 $\pm$ 0.4 &\textbf{94.4 $\pm$ 0.3} \\
Kag 3& \worse \textit{93.2 $\pm$ 0.6} & \worse 80.0 $\pm$ 0.4 & \worse 90.2  $\pm$ 0.6 &\textbf{94.3 $\pm$ 0.3} \\
Kag 4& \better \textbf{94.7} $\pm$ 0.5 & \worse 80.2 $\pm$ 0.3 & \worse 88.8 $\pm$ 0.5  &\textit{94.1 $\pm$ 0.2} \\
Kag 5& \worse \textit{92.8} $\pm$ 0.6 & \worse 77.7 $\pm$ 0.4 &\worse 87.7 $\pm$ 0.6 &\textbf{95.2 $\pm$ 0.2} \\
Pck 1& \better \textbf{90.5 $\pm$ 0.5} & \worse 71.3 $\pm$ 0.8 & \worse 82.4 $\pm$ 0.8 & \textit{87.2 $\pm$ 0.5}\\
Pck 2& \better \textbf{88.6 $\pm$ 0.5} & \worse 71.3 $\pm$ 0.2 & \worse 80.6 $\pm$ 1.3 &\textit{86.2 $\pm$ 0.5} \\
Pck 3&\,\,\, \textbf{88.4 $\pm$ 0.5} & \worse 71.5 $\pm$ 0.4 & \worse 82.8  $\pm$ 0.6 &\textbf{88.4 $\pm$ 0.4} \\
Mixed& \better \textbf{93.5 $\pm$ 0.5} & \worse 73.9 $\pm$ 0.3 & \worse 85.4  $\pm$ 0.7 &\textit{89.1 $\pm$ 0.4} \\
\end{tabular}
\caption{Accuracy Results for all datasets and techniques. The best results are remarked in bold. The second best results are remarked in italic. The \better and \worse symbols indicate whether a technique is statistically better or worse to EnTS respectively, according to Wilcoxon test.}
\label{tab:acc}
\end{table}

Table \ref{tab:acc} shows the direct comparison between the four techniques discriminating malware and benign-ware, according to the accuracy. It divides the results by technique and provides the accuracy of applying each algorithm to the specific datasets described in section \ref{sec:ExpSep:data}. For Kaggle data, EnTS and NCD generally obtain the best results (EnTS is over 94\% of accuracy in all cases and NCD over 92\%). SE is always worse than these techniques and CR is the worst approach. For VirusShare or packed data, all techniques reduce their accuracy, however, EnTS and NCD keeps competitive results compare with the rest of the techniques (over 86\% and 88\% in all cases). These results might be affected by similar packing systems used by malware and benign-ware. The mixed data shows that NCD obtains the best results followed by EnTS (93.5\% and 89.1\% of accuracy, respectively). SE obtains similar results than in the previous case and CR is the worst technique. This analysis shows that EnTS and NCD are the best techniques classifying malware. 


\begin{table}[t]
\centering
\footnotesize

\begin{tabular}{l|r||rrrrr}
 & 0 & 0.002 & 0.01 &0.05 & 0.1 & 0.15\\
\hline
Kaggle 1 NCD & \textit{0.72}&0.72 & 0.73 & 0.98 & 0.99 & 1.00\\
Kaggle 1 CR & 0.23& 0.30& 0.32 &0.50 & 0.61 & 0.72\\
Kaggle 1 SE & 0.42&0.52& 0.66 & 0.81 & 0.86 & 0.90\\
Kaggle 1 EnTS & \textbf{0.78}&0.85&0.91 & 0.96 & 0.98 & 0.99\\
\hline
Kaggle 2 NCD & \textit{0.69} &0.69& 0.77 &0.98 & 0.99 & 0.99\\
Kaggle 2 CR & 0.30& 0.33& 0.40 & 0.53 & 0.64 & 0.73\\
Kaggle 2 SE & 0.44 & 0.45 & 0.58 & 0.82 & 0.88 & 0.89\\
Kaggle 2 EnTS & \textbf{0.80} &0.80 & 0.88 & 0.95 & 0.97 & 0.98\\
\hline
Kaggle 3 NCD & \textit{0.70}& 0.72& 0.81 &0.97 & 0.99 & 0.99\\
Kaggle 3 CR & 0.27& 0.32& 0.34 & 0.58 & 0.68 & 0.74\\
Kaggle 3 SE & 0.42 &0.56 &0.65 & 0.84 & 0.91 & 0.93\\
Kaggle 3 EnTS & \textbf{0.78} &0.84& 0.90 & 0.95 & 0.96 & 0.98\\
\hline 
Kaggle 4 NCD & \textit{0.73} & 0.76 & 0.87 & 0.99 & 1.00 & 1.00\\
Kaggle 4 CR & 0.26 & 0.30& 0.35 & 0.53 & 0.65 & 0.76\\
Kaggle 4 SE & 0.42 & 0.51 & 0.68 & 0.82 & 0.88 & 0.91\\
Kaggle 4 EnTS & \textbf{0.77} & 0.80 &0.88 & 0.94 & 0.96 & 0.98\\
\hline
Kaggle 5 NCD & \textbf{0.79}& 0.82&0.91 &0.95 & 0.98 & 0.99\\
Kaggle 5 CR & 0.25 & 0.26& 0.35 & 0.51 & 0.65 & 0.71\\
Kaggle 5 SE &0.36 & 0.41 & 0.64 & 0.82 & 0.84 & 0.88\\
Kaggle 5 EnTS & \textbf{0.79} & 0.82 & 0.91 & 0.95 & 0.98 & 0.99\\
\hline
Packed 1 NCD & \textit{0.10}&0.10 & 0.23 &0.92 & 0.96 & 0.96\\
Packed 1 CR & 0.00& 0.00& 0.00 & 0.27 & 0.53 & 0.56\\
Packed 1 SE & 0.09&0.09& 0.44 & 0.69 & 0.77 & 0.80\\
Packed 1 EnTS & \textbf{0.26}&0.26&0.44& 0.76 & 0.83 & 0.90\\
\hline
Packed 2 NCD & \textit{0.06} &0.06& 0.11&0.84 & 0.88 & 0.92\\
Packed 2 CR & 0.00& 0.00& 0.08 & 0.30 & 0.53 & 0.61\\
Packed 2 SE & 0.11 & 0.11 & 0.35 & 0.66 & 0.74 & 0.75\\
Packed 2 EnTS & \textbf{0.26} &0.26 & 0.44 & 0.69 & 0.80 & 0.92\\
\hline
Packed 3 NCD & \textit{0.13}& 0.13& 0.29 &0.83 & 0.89 & 0.92\\
Packed 3 CR & 0.00& 0.00& 0.09 & 0.27 & 0.53 & 0.62\\
Packed 3 SE & 0.11 &0.11 &0.31 & 0.68 & 0.74 & 0.82\\
Packed 3 EnTS & \textbf{0.28} &0.28& 0.44 & 0.75 & 0.87 & 0.92\\
\hline
Mixed NCD & \textbf{0.52} & 0.60 & 0.82& 0.92& 0.95& 0.98\\
Mixed CR & 0.17& 0.18& 0.23 & 0.42 & 0.53 & 0.61\\
Mixed SE & 0.11 &0.26 &0.52 & 0.78 & 0.82 & 0.85\\
Mixed EnTS & \textit{0.35} &0.45& 0.69 & 0.82 & 0.88 & 0.91\\

\end{tabular}
\caption{False positives and true positives rates for all techniques and datasets. The ROC curve that has been chosen is the median of all the ROC curves generated during the experimental process.}
\label{tab:fp}
\end{table}

Increment precision is equivalent to reduce false positives. The ml classifier penalizes false positives during the learning process, as was mention above, to ensure that the model generates a good diagnosis of the malicious programs. The cut-off or threshold used in the ROC curve also provides a confidence value in the random forest voting system that helps to reduce false positives. Using 10 cross-fold validation in the training set, we set the cut-off to the most conservative value, i.e. the one that ensures 0 false positives in all validation sets, and we apply this last model to the test data. 

Table \ref{tab:fp} shows the median results for the ROC curves for all the experiments. In this table, we can see how the threshold variation modifies the true/false positives rates for each dataset. For Kaggle data, EnTS detects more than the 77\% of malware with 100\% precision (i.e. 0 false positive) and NCD detects 69\%. SE only detects, at most, 44\%. For packed malware, EnTS detection rate is reduced to 26-28\%, NCD to 6-13\%, SE to 11\% and CR to 0\%. For mixed data, NCD improves its results significantly (52\%), as well as EnTS (35\%). CR is also incremented (17\%). This table has shown that EnTS outperforms all techniques with 100\% precision. These results also discard CR as a classifier. After setting the threshold to the 100\% precision, the median accuracy achieved by EnTS for Kaggle data is 89\% $\pm$ 5.7, for packed data is 63\% $\pm$ 5.8  and for mixed data is 66.2\% $\pm$ 2.1. The mixed data model will be used for ITect.

Finally, we want to go deeper in the specific concealment strategies used by the families and packing systems and how they affect the performance of each technique.
\begin{table}
\centering
\footnotesize

\begin{tabular}{l|rrrr}
\multicolumn{1}{c|}{Class}& \multicolumn{1}{c}{NCD} & \multicolumn{1}{c}{CR} & \multicolumn{1}{c}{SE}& \multicolumn{1}{c}{EnTS}\\
\hline
Armadillo &\better \textbf{95.9}$\pm$1.9  &\worse 82.4$\pm$2.9  &\worse 85.1$\pm$2.5  &\textit{89.2}$\pm$3.5\\
ASPack &\better 85.7$\pm$11&\textit{42.9}$\pm$15  &\textit{42.9}$\pm$7.3  &\textit{42.9}$\pm$14\\ ASProtect &\better \textbf{100}$\pm$1.9&\worse 95.2$\pm$2.4  &\textit{95.8}$\pm$3.1  &\textit{95.8}$\pm$2.5\\
Borland &\textbf{100}$\pm$0.0&\textbf{100}$\pm$5.3 &\textbf{100}$\pm$2.3 &\textbf{100}$\pm$3.7\\
NET &\better  \textbf{100}$\pm$0.9&\worse 53.8$\pm$5.1  &\worse 73.1$\pm$3.0  &\textit{96.2}$\pm$3.1\\
PEComp &\better  \textbf{83.9}$\pm$11&\worse 45.2$\pm$11  &\textit{50.0}$\pm$5.6  &\textit{50.0}$\pm$7.5\\
UPX &\better \textbf{100}$\pm$1.1&95.2$\pm$3.9  &\worse 91.3$\pm$3.5  &\textit{95.7}$\pm$2.2\\
Rest &\better \textbf{97.7}$\pm$1.5&\worse 83.7$\pm$6.3  &\textit{93.0}$\pm$3.2  &92.3$\pm$4.4\\

     \hline
Ramnit &\worse 57.9$\pm$6.4& \worse  \textit{71.8}$\pm$4.5& \worse 52.1$\pm$5.2&  \textbf{84.7}$\pm$4.2\\
Lollipop & \textbf{97.1}$\pm$1.4& \worse  55.3$\pm$5.9& \worse  84.8$\pm$2.0&  \textit{97.0}$\pm$1.2\\
Kelihos3 &\textbf{100}$\pm$0.2& 99.4$\pm$1.1& \textbf{100}$\pm$0.2& \textbf{100}$\pm$0.0\\
Vundo &\worse 72.4$\pm$12& \worse 86.2$\pm$8.0& \textbf{100}$\pm$2.7& \textbf{100}$\pm$1.6\\
Simda &\textbf{100}$\pm$0.0& \worse 50.0$\pm$38& \worse  75.0$\pm$22& \textbf{100}$\pm$4.0\\
Tracur &\worse  \textit{95.7}$\pm$2.6&\worse 90.0$\pm$5.9& \worse  93.5$\pm$4.6& \textbf{100}$\pm$1.1\\
Kelihos1 &\better \textbf{96.0}$\pm$4.2&\worse  91.7$\pm$11&   \better \textbf{96.0}$\pm$4.0&  95.0$\pm$5.0\\
Obf.ACY &\worse \textit{80.8}$\pm$4.3& \worse 53.2$\pm$11& \worse  68.5$\pm$4.5&  \textbf{91.9}$\pm$2.2\\
Gatak &\worse 91.3$\pm$4.1& \worse 83.1$\pm$5.0&  \better \textbf{98.6}$\pm$1.3&  \textit{94.6}$\pm$2.4\\

\hline
Poly &\better \textbf{97.3}$\pm$0.6& \worse 74.2$\pm$0.6&\worse  83.4$\pm$1.1 &\textit{93.3}$\pm$0.5\\
Meta &\worse \textit{94.1}$\pm$0.8& \worse 80.1$\pm$0.6 &\worse  86.4$\pm$1.0& \textbf{94.6}$\pm$0.7\\
Packed &\better \textbf{80.2}$\pm$2.6& \worse 67.1$\pm$0.6 & \better \textit{75.7}$\pm$1.0 &73.4$\pm$1.1\\

\end{tabular}
\caption{Breakdown of Table \ref{tab:acc} results by: malware families in Kaggle dataset, packing systems in packed dataset and concealment strategy in mixed dataset.}
\label{tab:mpp}
\end{table}
Table \ref{tab:mpp} breakdowns the results from Table \ref{tab:acc} in families, packing systems and strategies for Kaggle, VirusShare and mixed data, respectively. For packed malware, NCD achieves the best performance in all cases, followed by EnTS. All techniques are good discriminating Borland system, as well as ASProtect and UPX. For Kaggle families, we can see that NCD and EnTS outperforms the rest of techniques in almost all cases. This also shows the effectiveness of EnTS when it is applied to metamorphic malware. Due to metamorphic malware has not intuitive entropy variations we focus on the two specific families: Vundo and Obfuscator.ACY. Vundo was previously studied by Li et al., who provided a description about the metamorphic engine \cite{li2011mechanisms}. This description mentions that the data section is encrypted or compressed, therefore this produces entropy variations that can be detected by EnTS. This fact is also detected in the entropy profile, where there are long sections with higher entropy than others. For Obfuscator.ACY the previous pattern is also frequent in the entropy profile, but in smaller sections, probably related to encrypted or compressed strings. These variation patterns make the metamorphic data totally unique for EnTS, and it is the reason it can easily detect them. For mixed data, the best results are for polymorphic and metamorphic data, applying NCD and EnTS. For packed, NCD is the best, followed, in this case, by SE which is close to EnTS. 

\hspace{-0.5cm}\fbox{\begin{minipage}{\columnwidth}
Research question 1 asks whether EnTS is accurate and precise.  It \emph{is} precise:  it obtains 100\% precision.  However, it falls short of the 90\% accuracy bar. It obtains 89\% accuracy on Kaggle, 66.2\% on mixed data and 63\% on packed. This motivates ITect, which combines SLaMM and EnTS.
\end{minipage}}

\hspace{-0.5cm}\fbox{\begin{minipage}{\columnwidth}
Research question 1.2 asks about comparing EnTS with the other techniques. EnTS is more accurate than CR and SE and similar to NCD. These results show that NCD and EnTS are competitive classifiers in all cases. CR does not provide any relevant information to the system and EnTS easily defeats SE. Besides, EnTS and NCD can handle specific families and packers, forcing malware writers to create new ones. 
\end{minipage}}

\paragraph{SLaMM evaluation}
Table~\ref{tab:acc2} reports the classification accuracies and false positive rates for SLaMM in terms of the three classifiers for different malware dataset, namely metamorphic, polymorphic and packed. For metamorphic and polymorphic malware, $C_D$ achieves the best accuracy, 94.7\% and 83.4\%, respectively. Whereas $C_{\mathit{MSE}}$ has the best detection accuracy for packed malware, 86.1\%. In terms of false positive rates, the performance of $C_X$ is notable as it has the lowest false positive rate for all three different malware datasets. The accuracy and false positive rates of SLaMM, highlighted in bold, depends on individual classifiers. Interestingly, SLaMM has the best accuracy for metamorphic malware and it has the lowest false positive rate for polymorphic. SLaMM achieves a relatively good average accuracy of 72\% with sufficiently low average false positive 0.027.

  \begin{table}[!h]  
  \centering
  \footnotesize
  \begin{tabular}{ p{.4cm} p{.3cm} p{.3cm} p{.3cm} p{.85cm} |p{.4cm} p{.4cm} p{.4cm} p{.85cm}}
    {\textbf{Cat.}}  & \multicolumn{4}{c|}{\textbf{Accuracy}}& \multicolumn{4}{c}{\textbf{False Positive}} \\
    
    & \text{$C_D$} & \text{$C_{MSE}$} & \text{$C_X$}&\textbf{SLaMM}& \text{$C_D$} & \text{$C_{MSE}$} & \text{$C_X$}&\textbf{SLaMM}\\
    \hline
     Meta &94.7&89.1 & 79.9 &\textbf{76.1}  &0.052 &0.013&0.01&\textbf{0.011}\\ 
    Poly & 83.4& 75.0&  70.1 &\textbf{67.2}&0.013 &0.002&0.001&\textbf{0.0}\\ 
    Pkd & 83.7 & 86.1 & 85.1 &\textbf{73.6}&0.118 &0.232 & 0.077&\textbf{0.07}\\ 
\hline
    Avg. & 88.2 & 83.4 & 78.4 & \textbf{72.3} &0.061& 0.082 & 0.029 & \textbf{0.027}\\ 
      \end{tabular}
  \vspace{.2cm}
  \caption{Detection Accuracy and False Positive Rate Comparison of 3 classifiers with different malware categories.}
   \label{tab:acc2}
\end{table}

SLaMM can use different configurations for the malware language models and, in order to improve the quality ratios, we have combined the three malware language models of previous sections using the classifiers. This combination helps to detect specific malware that can be detected by a language model and not for other. For example, for metamorphic malware, the opaque invariants would be detected by the metamorphic engine and for packed malware, specific details like the 3-gram signature of UPX packer would be detected by the packed language model. The combination is performed as follows: for reconstruct each classifier as the OR of the equivalent classifiers in each language model. After, we perform the AND operation among the different classifiers.
Using this new configuration, SLaMM achieves 82.1\% of accuracy with 0 FP for the mixed dataset. 

\hspace{-0.5cm}\fbox{\begin{minipage}{\columnwidth}
Research question 2 asks whether SLaMM is accurate and precise. It \emph{is} precise: it obtains 100\% precision. However, it fails in the accuracy. The accuracy of SLaMM is 82.1\% on the mixed data. This also motivates ITect. 

\end{minipage}}

\paragraph{Scalability}

Table \ref{tab:time} shows the average time consumption of the techniques for training and testing. The table is divided in three datasets (Kaggle, packed and mixed), and three specific values: the space generation or training (where the algorithms generate the similarity matrices, entropy profiles or the compressibility values), the classification process and the total average time. EnTS outperforms every single technique. We can also see that NCD is the most impractical technique, taking 2 days in the best case and 5 in the worse. This shows that NCD is not optimal for malware detection. It is a consequence of the file compression and the pairwise comparison to generate the similarity matrix. The compression process also affects to CR which needs more time to calculate the ratios. The pairwise comparison affects to NCD and SE. EnTS uses no pairwise comparison, and this improves the time consumption. Besides, the entropy profile generation and the wavelet decomposition are linear processes, they do not generate a bottleneck during the analysis. 

\begin{table}
\centering
\footnotesize

\begin{tabular}{l|rrrr}
 & \multicolumn{1}{c}{NCD} & \multicolumn{1}{c}{CR}& \multicolumn{1}{c}{SE} & \multicolumn{1}{c}{EnTS}\\
\hline
Kaggle Space Gen & $>$5days & 60m & 140m & 5m\\
Kaggle Classification & 320m & 3m & 300m & 4m\\
Kaggle Total & $>$5days & 63m & 440m & \textbf{9m}\\
\hline
Packed Space Gen & $>$2days & 40m & 90m & 3m\\
Packed Classifications & 31m &1m & 32m & 2m\\
Packed Total & $>$2days & 	41m & 122m & \textbf{5m}\\
\hline
Mixed Space Gen & $>$5days & 60m & 140m & 5m\\
Mixed Classification &320m & 3m & 300m & 4m\\
Mixed Total & $>$5days & 63m & 440m & \textbf{9m}\\
\end{tabular}
\caption{Average time results for the different methods and all the databases. Time is approximated in minutes (m) and days.}
\label{tab:time}
\end{table}

The memory consumption of each metric grows depending on the space size. For NCD and SE, this space is related to the similarity matrix, which grows as $O(P^2)$ while EnTS grows linearly $O(P)$ according to the number of programs, $P$, due to the number of coefficients (or coordinates) used in the space is fixed. CR also grows linearly according to the number of files. 

Table \ref{tab:ng} shows computation time and memory consumption for SLaMM. The time consumption depends on two factors, the order of $n$-gram language model and the size of the corpus in terms of individual file size. The model size increases exponentially when the $n$-gram order becomes higher. However, models can be generated in parallel. Therefore, the time of the whole system corresponds to the longest language model generation process. For instance, SLaMM finishes the whole classification process including, pre-processing, $3$-gram LM construction for each malware dataset, LM look up and detection in 93 minutes (detection takes 23 minutes using all classifiers). On the other hand the memory consumption does not depend of the corpus size or individual file size in the corpus. It only depends on the order of the language model. SLaMM occupied 273M of RAM during classification and it remains unchanged for each dataset. 

The time consumption ranking for the techniques and for datasets containing 2000 malware and 2000 benign-ware starts with NCD consuming more than five days. It follows with SE consuming 440 minutes, SLaMM consuming 116 minutes, CR consuming 63 minutes and finally EnTS consuming only 9 minutes.
The equivalent memory consumption ranking starts with NCD and SE consuming consuming a big square similarity matrix ($O(P^2)$). It follows with SLaMM (depending on the value of $n$ in $n$-grams) and EnTS and CR as $O(P)$ techniques.

  \begin{table}[!h]  
  \centering
  \footnotesize
  \begin{tabular}{ l | r | r }
   \multicolumn{1}{c|}{Malware} &  \multicolumn{1}{c|}{Time} & \multicolumn{1}{c}{Memory}\\
   \hline
    Metamorphic &53m & 273M\\ 
    Polymorphic & 71m&273M\\ 
    Packed & 45m &273M \\ 
    Benign &93m & 273M \\ 
    
  \end{tabular}
  \vspace{.2cm}
  
 \caption{Time and memory consumption of SLaMM with $3$-gram LM}
   \label{tab:ng}
\end{table}

\hspace{-0.5cm}\fbox{\begin{minipage}{\columnwidth}
Research question 3 asks whether SLaMM and EnTS scale better than NCD, SE and CR. They \emph{do} scale better and are linear scalable, but in the case of SLaMM compared with CR. 
\end{minipage}}

\paragraph{ITect Evaluation}
\label{subsect:joining}

In order to evaluate how EnTS and SLaMM complement each other for classification, we have chosen the mixed dataset, which is the most heterogeneous. The combination is a simple OR, named ITect. The test results for the mixed dataset using ITect are: 90.3\% $\pm$ 0.79 of accuracy and 0 false positives.
In real environments, it is more frequent to find benign-ware than malware, then we have designed an experiment changing the balance between them. We set the number of malware instances as a percentage starting from 0\% to 50\%. We apply ITect to measure the false positives rate and the accuracy. 

\begin{figure}
\centering
\includegraphics[width=0.5\textwidth]{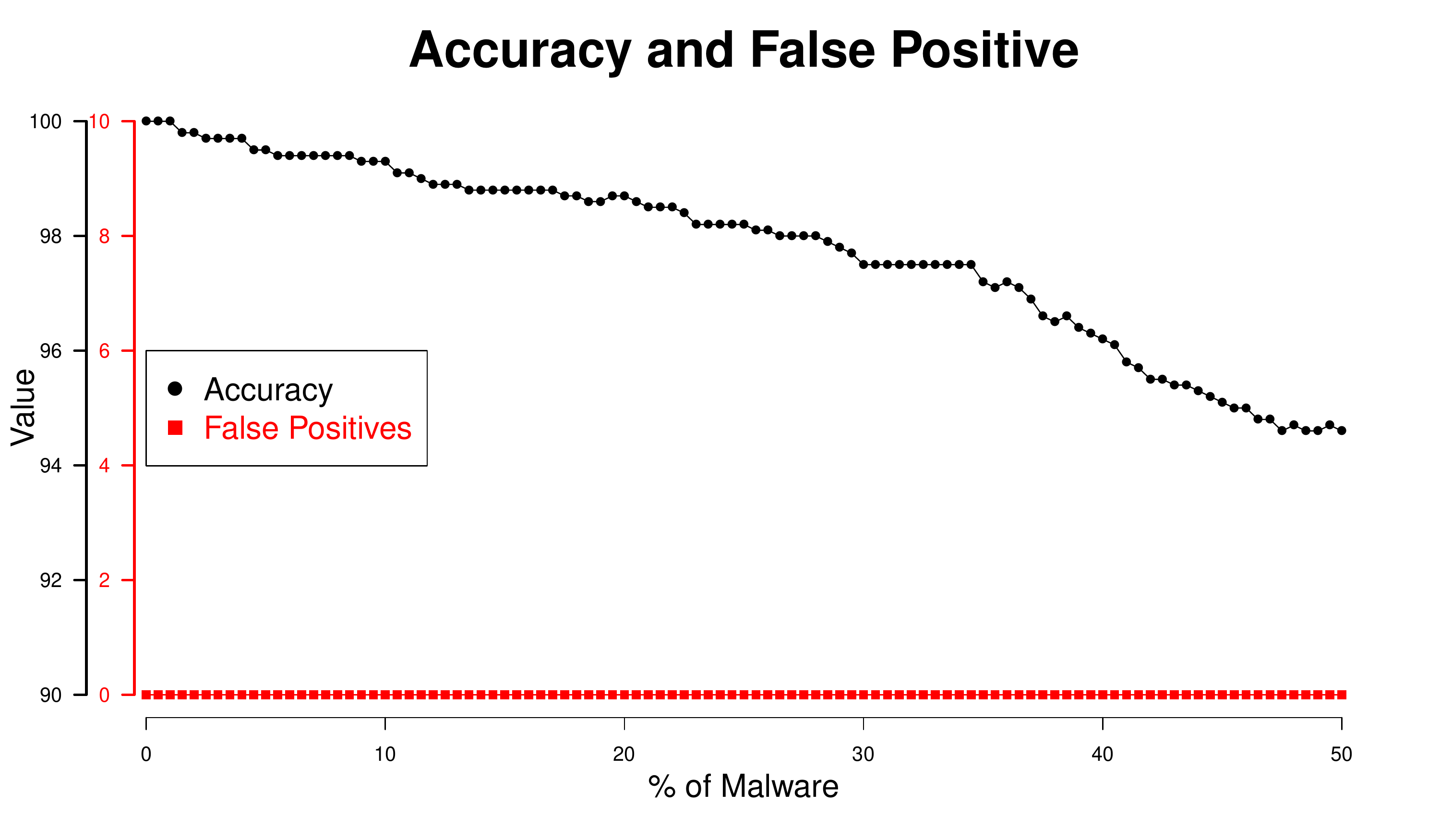}
\caption{Accuracy and False Positive Rate for ITect. This is the median results for 100 experiments modifying the percentages of malware from 0 to 50\% of the total data.}
\label{fig:finalResult}
\end{figure}

Figure \ref{fig:finalResult} shows the results. The precision keeps constant in 100\% (as the false positives are 0), therefore no benign-ware is detected as malware. The accuracy starts with a 100\%, and it goes down when some malware is not detected (false negatives). It linearly goes down until achieving 94.6\% of accuracy with 50\% of malware programs.

To provide an intuition about recall instead of precision, it is easy to deduce that it goes down with the accuracy. The accuracy is defined by $(TP + TN)/(TP+TN+FP+FN)$, then, when the $FP$ is constant and 0, it means that the accuracy and the $FN$ are inversely proportional. Due to recall depends on the $TP$ and $FN$, it means it goes down. However, during the construction process, we might change the penalization factor from $FP$ to $FN$, if we want to increment the recall (which might be interesting for other kind of analysis), but reduce both values (i.e. $FP$ and $FN$) supposes to increment the accuracy, which is more challenging.

We have also perform a sanity check using benign-ware and malware downloaded from download.com and VirusShare, respectively, dated on January 2016. We have download 10 malware and 10 benign-ware. All techniques were able to identify all the benign-ware. EnTS identified 5 of the malware files while SLaMM identified 6. ITect identified 8 malware files with 100\% precision. 

\hspace{-0.5cm}\fbox{\begin{minipage}{\columnwidth}
Research question 4 asks whether ITect is accurate and precise. It \emph{is} accurate and precise. ITect classifies malware with 90.3\% of accuracy and 100\% of precision, providing a competitive wait to detect malware. 
\end{minipage}}

\paragraph{Comparing ITect with AV Engines}

We have compared ITect with 56 Anti-Virus Engines. For this comparison, we have sent all the test set from the mixed dataset to Virus Total. In the case of packed malware, all the data was already classified as Malware using this system, but Kaggle data is fresher and there are a few anti-virus that can detect it. Table \ref{tab:avcomp} shows the comparison between the best and worst engines related to accuracy and precision. We can see that SLaMM and EnTS obtain the best accuracy results with 100\% precision. Avast accuracy is higher than EnTS, but its is penalized by the false positives. ITect outperforms all the system achieving an accuracy of more than 90\% with 100\% precision.

\begin{table}
\footnotesize

\centering
\begin{tabular}{c|c|r|r}
& Technique & \multicolumn{1}{c|}{ACC} & \multicolumn{1}{c}{Precision}\\
\hline
Best AV with 0 FP& ESET-NOD32 & 65.6\%& 100.0\%\\
More Accurate AV & Avast & 71.4\%& 99.4\%\\
AV with higher FPs&Emsisoft & 64.9\%&97.2\%\\
AV with worst Accuracy&Zoner& 50.8\%&100.0\%\\
&SLaMM & 82.1\%&100.0\%\\
&EnTS&66.2\%&100.0\%\\
&ITect& \textbf{90.3}\% & \textbf{100.0}\%\\
\end{tabular}
\caption{Comparison between Anti-Virus Engines, EnTS, SLaMM and ITect, according to precision and accuracy.}
\label{tab:avcomp}
\end{table}

\hspace{-0.5cm}\fbox{\begin{minipage}{\columnwidth}
Research question 5 asks about ITect performance compared with the 56 AV Engines from VirusTotal. ITect \emph{outperforms} all AV Engines with 100\% of precision. Besides, it is almost 20 points higher than the most accurate AV Engine. 
\end{minipage}}

%% file: countermeasures.tex
\section{Countermeasures}
\label{sect:counter}
As a move (or moves) in the arms race we need to consider possible countermeasures. Our moves are useful if it is \emph{difficult} for the opponents to counter them, even if it is possible to counter them.

\paragraph{EnTS}

First consider EnTS. Opponents can attempt to modify polymorphic malware so that the entropy profile is closer to that of benign programs. The evidence from our experiments is that benign programs typically have an entropy profile which is smooth, i.e. with low variation, and which has an average entropy which is lower than that of polymorphic malware.

Clearly, to evade EnTS, opponents need to ensure that the average entropy of an EnTS chunk is lower and that the overall variation is lower. To achieve this they need to add in low entropy material within each chunk. (Adding material in amounts longer than a chunk can reduce the average variability but fails to evade EnTS as it simply perpetuates the existing malware entropy profile by adding more low entropy regions that contrast with the existing high entropy ones). Adding low entropy material to each high entropy chunk will definitely have a smoothing effect on the variability but it is problematic with respect to achieving an average entropy over the chunks which is similar to the average entropy of benign-ware. To bring down the overall chunk average the entropy of the added material must be lower than the average chunk entropy of the  benign-ware. The risk to the opponent is that it is not possible to do this and still look like benign code. The measure of the opponent's ability to do this is in some sense the measure of any gap between the representativeness of the benign-ware zoo used to construct the average chunk entropy and benign-ware at large.

Let $O$ be the average entropy of the material that the opponent needs to add, $M$ be the average entropy of the benign-ware, and $N$ be the average entropy of the polymorphic malware. Then $O \le M \le N$. Consider a chunk and let $n$ be the number of chunks of low entropy material that need to be added to the malware to achieve an average chunk entropy of $M$. Then for a single high entropy chunk the opponent requires
\begin{align*}
\frac{N+nO}{n+1} =&\ M, \mbox{ i.e. } n = \frac{N-M}{M-O}
\end{align*}
The number, $n$, of additional chunks worth of material to be added for each chunk of packed/encrypted material is given by the ratio $\frac{N-M}{M-O}$. Clearly when $O$ is close to $M$ (and we have argued above that it must be) this number is very large, a threat to the viability of the malware, and the desired outcome is impossible to achieve in the case that the benign zoo is highly representative of the entire benign program population -- as then $O$ must be very close to $M$ and the amount of material to add becomes infeasible.

\paragraph{SLaMM}

\begin{figure}
\centering

\usetikzlibrary{shapes,backgrounds}
\def\firstcircle{(0,0) circle (1.2cm)}
\def\secondcircle{(0:2cm) circle (1.2cm)}
\def\thirdcircle{(0:2cm) circle (1.2cm)}

\begin{tikzpicture}
\draw (-3,-1.3) -- (5,-1.3) -- (5,1.3) -- (-3,1.3) -- (-3,-1.3);
\draw (-2,0.1) node[align=center]{Semantically\\ equilavent\\ to M};
\draw (4.2,0) node[align=center]{Statistical\\ profiled\\ benign\\ programs};
\draw (1,1.5) node{Programs};
   \draw \firstcircle node {$\cdot M$};
    \draw \secondcircle node {};
   \begin{scope}
      \clip \firstcircle;
      \fill[red] \secondcircle;
    \end{scope}

 \end{tikzpicture}
 \caption{Semantic space for malware and benign-ware.}
  \label{Shrunken}
\end{figure}
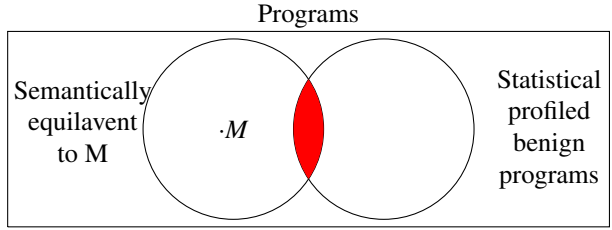

Now consider SLaMM. The strength of SLaMM is that we construct a highly accurate, fine-grained, statistical model of benign-ware and use information theory to measure divergence from that model relative to a malware model. In order to evade detection it is no longer enough for the opponent to metamorphically produce a semantically equivalent malware. It now must also have the statistical profile of benign programs.  The first point to make is that SLaMM shrinks the available space into which any program can be metamorphosed, as illustrated in figure \ref{Shrunken}

The second point to make is that guaranteeing the statistical profile of the binary is a novel and non-trivial exercise. Imagine the opponent writing malware source code from scratch. There is the crap-shoot approach. If it is sufficiently novel and different to both existing models it may fortunately be more different to the malware zoo than to the benign-ware zoo. Or there is the benign model aware approach. She needs an IDE with a built-in interpreter and an analyser that constantly checks the statistical profile of the low-level code. The editor perhaps starts with the source code blue for ``cool'' or ``OK'' and as the low-level code deviates from the required statistical profile, gradually turns the colour to red. As there is a level of indirection and no obvious guidelines, once the code turns red it is not clear what to do. Delete recent sections? start again?

Alternatively, she could turn to search-based software engineering and use genetic program improvement (GI) to search for an equivalent program to an existing malware with the right statistical profile \cite{langdon2013optimising}. As argued above we have shrunken the space of equivalent programs. Then, maintaining program equivalence is difficult in this arena unless she restricts the search space to that defined by semantics preserving local transformations, an approach deprecated in the GI community as too restrictive. In addition, although it may be straightforward to define a fitness function, it is not clear until someone tries this in anger that the resulting fitness landscape has a gradient that allows the search to succeed.

%% file: relatedWork.tex
\section{Related Work}

We touch upon the problems with the state of the art in malware analysis, the growing usefulness of network association metrics, and then review byte level content analysis as a detection technique. 

\paragraph{Detection via Byte Level Content}
In 1994 Kephart presented an n-gram approach for extracting signatures but reported no results ~\cite{Kephart94}. In 2001,  Schultz et al. used several data mining techniques on binaries to distinguish between benign and malicious executables in Windows or MS-DOS format.  Memory consumption was a scalability bottleneck. They experimented on a dataset of 3265 malware and 1001 benign files but lacked fresh data for testing. Validation achieved 97.11\% Accuracy with 3.8\% FP rate ~\cite{SchultzEZS01}.

In 2006, Kolter and Maloof  used Information Gain combined with byte level analysis of n-grams to classify and detect malware. Again they did not use fresh data for the test phase. They experimented on two small datasets, one of 476 malware, 561 benign-ware (95\% accuracy with 5\% FP in validation); the second of 1971 benign-ware, 1651 malware (94\% accuracy and 1\% FP in validation) \cite{KolterM06}.

In 2007, Lyda and Hamrock used the average entropy of a whole file and the entropy of specific code sections (discovered only by using static analysis). They showed that binary files with a higher entropy score tend to be correlated with the presence of encryption or compression.  They compared more than 20,000 malware to check whether they are able to detect these concealment methods but did not consider malware detection~\cite{LydaH07}. In the same year, Stolfo et al used 1-gram and 2-gram byte distributions for a file to compare it with different filetype models for filetype identification~\cite{Li05} in malware detection within DOC and PDF files. They reported on experiments with over 140 pdfs and 361 benign and 616 malware and results with between 3\% to 20\% of false positives but no accuracy information. This work considered n-grams in a vector space, using their frequency and variation as features, but each dimension was a n-gram resulting in exponential increase in the number of dimensions~\cite{StolfoWL07}.

Tabish et al. in 2009 divided files into blocks, and calculated frequency of n-gram histograms for each block, then extracted statistical and information-theoretic features from the histogram to define a feature vector per block. They used this to classify a feature vector as normal or potentially malicious. Pairwise comparison between blocks of different files reduces the scalability of this approach. They claimed an accuracy rate of 90\% with a False Positive rate of around 10\%  \cite{TabishSF09}.

%
Santos et al. in 2011 introduced a semi-supervised methodology to reduce the labelling process. Their n-gram vector was the frequency of all possible n-grams, an important scalability limitation. After experiments on 1000 malware and benign-ware, they reported 89\% of accuracy with 10\% of false positives  \cite{santos2011semi}. 

Our work work has three advantages over previous work in detection via byte level content: (1) better accuracy combined with lower false positive rates, (2) better (linear) scalability in the detection phase, and (3) a more rigorous experimental approach.

\paragraph{Other Detection Methods}
Windows malware has become increasingly sophisticated at hiding itself and resisting analysis. Android malware currently lags behind this level of sophistication. 

Static analysis, whether based on abstractions of Control Flow Graphs and program semantics~\cite{PredaCJD07} or on opcode analysis~\cite{SantosBNPSLB10}, or focused on PE Headers and Static API Calls~\cite{XuSML07,YeLJW10} as features for machine learning, faces the increasing difficulty of initial reverse engineering.  In addition, Moser et al. demonstrated hard limits to the ability of static anaysis to deal with obfuscation~\cite{moser2007limits}. Dynamic analysis via virtual machines and sandboxes can avoid anti-disassembly measures but suffer from resistance via dynamic defence predicates and red pill environment detection techniques \cite{paleari2009fistful}. 
Windows malware analysis aiming to integrate dynamic and static analysis, as ~\cite{SantosDBNB12}~\cite{IslamTBV13}~\cite{salim15}, to produce features for data mining approaches suffer the same problems.

Recent approaches to Android malware exploit the relative lack of sophistication of that type of malware. These include Drebin~\cite{arp2014drebin},  CopperDroid~\cite{tam2015copperdroid}, which combine machine learning with behavioural models. Other tools as DroidSIFT~\cite{zhang2014semantics} are focused on anomaly detection and malware family classification. 

Malware detection tools focused on network neighbourhoods, for example, Nazca~\cite{invernizzi2014nazca} and AESOP \cite{Tamersoy:2014} 
show real promise in terms of scale and accuracy but require ground truth as a seed, just as our similarity techniques do.



%% file: conclusions.tex
\section{Conclusions}

ITect opens a new front in the arms race. Its level of abstraction makes it difficult to counter and it offers scalability advantages. We have demonstrated excellent precision and accuracy on a representative mixture of malware types drawn from the Kaggle malware data and VirusShare. We have demonstrated that both of its constituent detectors, EnTS and $\mathit{SLaMM}$, outperform previous information theoretic similarity measures. Indeed, ITect outperforms existing AntiVirus engines (as represented in VirusTotal) for accuracy and precision. Its time complexity is bounded above by the number of files to be classified. As an automated, execution agnostic, string-based similarity metric it offers wider scalability advantages beyond its time complexity class alone -- reducing human effort and reducing the need for dynamic or static analysis. 

%